\documentclass[aps]{revtex4}

\usepackage{graphics,epsfig}
\usepackage{amsmath}
\usepackage{amssymb}
\usepackage{amsfonts}
\usepackage{rotate}
\usepackage{epsfig}
\newcommand \be  {\begin{equation}}
\newcommand \bea {\begin{eqnarray} \nonumber }
\newcommand \ee  {\end{equation}}
\newcommand \eea {\end{eqnarray}}

\begin{document}
\title{Of Songs and Men: a Model for Multiple Choice with Herding}
\author{Christian Borghesi$^1$, Jean-Philippe Bouchaud$^{1,2}$}
\email{christian.borghesi@cea.fr;jean-philippe.bouchaud@cea.fr}
\affiliation{
$^{1}$ Service de Physique de l'{\'E}tat Condens{\'e},
Orme des Merisiers,
CEA Saclay, 91191 Gif sur Yvette Cedex, France.\\
$^2$ Science \& Finance, Capital Fund Management, 6-8 Bd
Haussmann, 75009 Paris, France.
}
\date{\today}

\newcommand{\nia}{n_i^\alpha}
\newcommand{\nib}{n_i^\beta}
\newcommand{\nja}{n_j^\alpha}
\newcommand{\njb}{n_j^\beta}
\newcommand{\fa}{F^\alpha}
\newcommand{\hia}{h_i^\alpha}
\newcommand{\aaa}{\alpha}
\newcommand{\bb}{\beta}
\newcommand{\ca}{{\cal C}^\alpha}
\newcommand{\ct}{{\cal C}}
\newcommand{\fia}{\phi^\alpha}
\newcommand{\nba}{d^\alpha}
\newcommand{\nbb}{d^\beta}
\newcommand{\ma}{m^\alpha}
\newcommand{\mb}{m^\beta}
\newcommand{\dmoy}{\langle d \rangle_i}

\begin{abstract} 
We propose a generic model for multiple choice situations in the presence of herding and compare it with
recent empirical results from a Web-based music market experiment. The model predicts a 
phase transition between a weak imitation phase and a strong imitation, `fashion' phase, where choices are 
driven by peer pressure and the ranking of individual preferences is strongly distorted at the aggregate level. 
The model can be calibrated to reproduce the main experimental results of Salganik et al. ({\it Science}, 
311, pp. 854-856 (2006)); we show in particular that
the value of the social influence parameter can be estimated from the data. In one of the experimental situation, this 
value is found to be close to the critical value of the model.
\end{abstract}

\maketitle

\section{Introduction}

Making decisions is part of everyday life. Some situations require a binary choice (i.e. to vote yes or no in a referendum,
to buy or not to buy a cell phone, to join or not to join a riot, etc. \cite{Schelling,Granovetter}). Many others involve multiple options, for 
example in the first round of French presidential elections (where the number of candidates is typically 15), in 
portfolio management where very many stocks are eligible, in
supermarkets where the number of possible products to buy is large, etc. In most cases, the choice is 
constrained by some generalized {\it budget constraint}, either strictly (at most one candidate in the French presidential 
election) or softly (the total spending in a supermarket should on average be smaller than some amount). It is common experience that people 
generally do not determine their action in isolation. Quite on the contrary, interactions and herding effects 
often strongly distort individual preferences, and are clearly responsible for the appearance of trends, fashions and 
bubbles that would be difficult to understand if agents were insensitive to the behaviour of their peers.
Catastrophic events (such as crashes, or sudden opinion shifts) can occur at the macro level, induced 
by imitation, whereas the aggregate behaviour of {\it independent} agents would be perfectly smooth. 

A relevant challenge in the present era of information economy is to be able to extract faithfully individual 
opinions/tastes from the publicly expressed preferences under the influence of the crowd. For example, book reviewers
on Amazon may be biased by the opinion expressed by previous reviews; if imitation effects are too strong, 
overwhelmingly positive (or negative) reviews cannot be trusted (see \cite{slanina}), as a result of 
``information cascades'' \cite{BHW}. In the case of financial markets, strong herding 
effects in the earning forecasts of financial analysts have been reported -- the dispersion of these forecasts is
typically ten time smaller than the {\it ex post} difference between the forecast and the actual earning (see \cite{guedj}
and refs. therein). These 
herding effects may lead to a complete divergence between the market price and any putative `rational' price. In the
context of scientific publications, the substitution of the present refereeing process by other assessment tools, 
such as number of downloads from a preprint web-page, or number of citations, is also prone to strong, winner-takes-all, 
distortions \cite{Redner,Russes}. More generally, it is plausible that such herding phenomena play a role in the appearance of Pareto-tails
in the measure of success (wealth, income, book sales, movie attendance, etc.).

Despite their importance, already stressed long-ago by Keynes and more recently by Schelling \cite{Schelling}, 
quantitative models of herding and interaction 
effects have only been explored, in different contexts, in a recent past, 
see \cite{Granovetter,Follmer,Galam,Kirman,Orlean,BHW,CB,MG,Matteo,MGreview}.
This category of models have in fact a long history in physics, where interaction is indeed at the root of genuinely 
collective effects in condensed matter, such as ferromagnetism, superconductivity, etc. One particular model, that appears 
to be particularly interesting and generic, is the so-called `Random Field Ising 
Model' ({\sc rfim}) \cite{Sethna}, which models the dynamics of magnets under the influence of a slowly evolving external 
solicitation. This model can be transposed in a socio-economics context \cite{Galam,QF,Nadal,QMB} to represent a 
{\it binary} decision situation under social pressure. A robust feature of the model
is that discontinuities appear in aggregate quantities when imitation effects 
exceed a certain threshold, even if the external solicitation varies smoothly with time. Below this threshold, 
the behaviour of demand, or of the average opinion, is smooth, but the natural trends can be substantially amplified 
by peer pressure. The predictions of the {\sc rfim} can be confronted, with some success, to empirical observations concerning sales of 
cell phones, birth rates and the terminal phase of clapping in concert halls \cite{QMB}. 

Here, we want to generalize the {\sc rfim} to {\it multiple} choice situations. One motivation is that, 
as mentioned above, these situations are extremely common. A more precise incentive for such a generalization 
is however the recent publication of a remarkable experimental paper by Salganik, Dodds and Watts \cite{exp}. In order to detect and
quantify social influence effects, the authors have conducted a careful Web-based experiment (described below) 
with several quantitative results. Their detailed interpretation begs for a specific model, which we introduce and discuss in 
this paper and compare with these empirical results. The model is found to fare quite well and allows one to extract from 
the data a quantitative estimate of the imitation strength, called $J$ below. Interestingly, one of the situations 
corresponds to a value of $J$ close to the critical point of the model, where collective effects become dominant 
and strongly distort individual preferences.

\section{The model}

We consider $N$ agents indexed by roman labels $i=1,...,N$, and $M$ items 
indexed by Greek labels $\alpha=1,...,M$. Each agent can construct his 
`shopping list' or portfolio of items, for simplicity, we restrict here to cases
where the quantity of item $\alpha$ is either zero or unity (in the example
of movies, we neglect the possibility of going twice to see the same movie). The 
portfolio of agent $i$ is therefore a vector of size $M$: $\{n_i^\alpha\}$ with 
$n_i^\alpha=0,1$. The ``budget constraint'' can in general be written as:
\be
B_{i}^- \leq \sum_{\alpha=1}^M n_i^\alpha \leq B_i^+,
\ee
where the budget might be different for different agents. 

The choices made by agent $i$ are assumed to be determined by three different factors:
\begin{itemize}
\item a piece of public information affecting all agents
equally, measuring the intrinsic attractivity of item $\alpha$. This is modeled by a real variable $F^\alpha$,
which may contain, for example, the price of the product (low price means large $F^\alpha$'s), or its technological 
performances, past reputation, etc. 
\item an idiosyncratic part describing the preferences/tastes of agent $i$, in the absence of any social pressure
or imitation effects. This part is again modeled by a real variable $h_i^\alpha$, which is positive and large 
if agent $i$ is particularly fond of item $\alpha$. 
\item a social pressure/imitation term which describes how the choices made by {\it others} affect the 
perception of item $\alpha$ by agent $i$. In full generality, we can write this term as:
\be
\sum_{j \neq i} \sum_{\beta} J_{j,i}^{\beta,\alpha} n_j^\beta
\ee
where $J_{j,i}^{\beta,\alpha}$ measures the influence of the consumption of product $\beta$ 
by agent $j$. Positive  $J_{j,i}^{\beta,\alpha}$'s  describe herding-like effects (which could
exist across different products), whereas negative $J_{j,i}^{\beta,\alpha}$'s are related to contrarian
effects (for example, agent $j$ buying item $\beta$ might push the price of item $\alpha$ up). 
We will consider in this paper a simplified version of the model where only the aggregate consumption 
of item $\alpha$ itself influences the value of $n_i^\alpha$, i.e.:
\be
J_{j,i}^{\beta,\alpha} = \frac{JM}{\cal C} \delta_{\alpha,\beta},
\ee
where the factor $M$ is introduced for convenience and $\cal C$ is the total consumption, defined as:
\be
{\cal C} = \sum_{i} \sum_{\alpha} n_i^\alpha.
\ee
\end{itemize}
We will also introduce the total consumption of item $\alpha$ as $\ca=\sum_i n_i^\alpha$, and the relative 
consumption (or success rate) $\phi^\alpha=\ca/\ct$, with $\sum_\alpha \phi^\alpha=1$.

We assume that the consumption of item $\alpha$ by agent $i$ is effective if the sum of these three determining factors 
exceed a certain {\it threshold}, and consider the following update rule for the $n_i^\alpha$'s:\footnote{We neglect 
$1/N$ corrections here.}
\be
n_i^\alpha(t+1) = \Theta\left[F^\alpha + h_i^\alpha + JM \left(\phi^\alpha(t) - 
\frac1M \right) - b_i(t)\right],
\ee
where $\Theta$ is the Heaviside function, $\Theta(u > 0)=1$ and $\Theta(u \leq 0)=0$. In the above equation, we have
added a `chemical potential' $b_i$ (borrowing from the statistical physics jargon) which allows the budget constraint 
to be satisfied at all times \cite{ustocome}. The $-1/M$ term was added for convenience, and makes explicit that it is the consumption of 
item $\alpha$ {\it in comparison with its expected average} $1/M$ that generates a signal (see also \cite{BG}). It is easy to
check that the case $M=1$, with $J \to J {\cal C}/N$, corresponds to the standard {\sc rfim} considered in \cite{QMB}. 
Note also that the $\Theta$ function describes a deterministic situation: as soon as the 
total `utility' of item $\alpha$ is positive for agent $i$, consumption is effective. One could choose a probabilistic
situation where $\Theta(u)$ is replaced by a smoothed step function, for example:
\be
\Theta_\beta(u)=\frac{1}{1+e^{-\beta u}}.
\ee
The limit $\beta \to \infty$ corresponds to the deterministic rule, to which we will restrict throughout this paper. 

In the following, we assume that both $F$'s and $h$'s are time independent, and 
taken from some statistical distributions which we have to specify. Here again, the number of possibilities is
very large, and correspond to different situations. We choose the $F^\alpha$'s as {\sc iid} random variables (for example
Gaussian), with mean $m_F$ and variance $\Sigma_F^2$. The mean $m_F$ describes the average intrinsic attractivity of 
items -- for example, a large overall inflation would lead to a negative $m_F$. The dispersion in quality of the 
different items is captured by $\Sigma_F$. More realistic models should include some sort of `sectorial' correlations
between the $F^\alpha$'s.

As for $\hia$'s, we posit that they can be decomposed as $\hia={\overline h}_i+\delta\hia$, where ${\overline h}_i$
describes the propensity of agent $i$ for consumption (`compulsive buyers' correspond to large positive 
${\overline h}_i$'s), whereas $\delta\hia$ correspond to the idiosyncratic tastes of agent $i$,
defined to have zero mean. For simplicity, we again assume that both $\overline h_i$'s and $\delta\hia$ are {\sc iid};
without loss of generality we can assume that the average (over $i$) of ${\overline h}_i$ is zero (a non zero value 
could be reabsorbed into $m_F$). The variance of ${\overline h}_i$ is $\Sigma^2$ and that of $\delta\hia$ is $\sigma^2$. 
Since in the limit $\beta \to \infty$ considered in this paper the overall scale of the fields is irrelevant, we can
choose to set $\sigma \equiv 1$. One could also add explicit time dependence, for example choosing $m_F$ to be an
increasing function of time, to describe a situation where the average propensity for consumption increases with time. 

The model as defined above is extremely rich and its detailed investigation as a function of the different parameters
and budget constraints will be reported in a forthcoming publication.  
The most interesting question about such a model is to know whether the realized consumption is {\it faithful}, i.e.
whether or not the actual choice of the different items reflects the `true' preferences of individual agents, as would be 
the case in the absence of interactions ($J=0$). Based on the {\sc rfim}, we expect that this will not be the case 
when $J$ is sufficiently large, in which case strong distortions will occur, meaning that the realized consumption
will (i) violate the natural ordering of individual preferences and (ii) become history dependent: a particular 
initial condition determines the `winners' in an irreproducible and unpredictable way. In order to characterize the inhomogeneity of 
choices, the authors of \cite{exp} have proposed and measured different observables, in particular:
\begin{itemize}

\item The Gini coefficient $G$, defined as:
\be
G = \frac{1}{2M} \sum_{\alpha,\beta} |\phi_\alpha-\phi_\beta|,
\ee
which is zero if all items are equally chosen, and equal to $1-1/M$ if a unique item is chosen. 
The Gini coefficient is a classic measure of inequality.
In fact, a more relevant measure of interaction effects is the ratio $G/G_0$, where $G_0$ is the Gini coefficient for $J=0$. 

\item The unpredictability coefficient $U$, defined as:
\be
U =
\frac{1}{M (\displaystyle{_2^W})} \sum_{\aaa=1}^M\sum_{k=1}^W \sum_{\ell<k}
|\fia_{(k)}-\fia_{(\ell)}|
\ee
where the indices $k, \ell$ refer to $W$ different `worlds', i.e. different realizations 
of the model {\it with the very same} $F^\alpha$'s but a different set $h_i^\alpha$'s (chosen with 
the same distribution) or different initial conditions. In the limit of a large population ($N \to \infty$), 
it is easy to show that $U=0$ when $J=0$, since the $\phi^\alpha$ only depends on the $F^\alpha$'s.
A non zero value of $U$, on the other hand, reveals that it impossible to infer from the intrinsic
quality of the items the aggregate consumption profile (strong distortion). 

\item A more detailed information is provided by the scatter plot of $\fia$ versus $\fia(J=0)$; for $J$ small one
expects a nearly linear relation, whereas for larger $J$ the points  acquire a larger dispersion 
and the average relation  becomes non-linear, indicating a substantial `exaggeration' of the 
consumption of slightly better items.

\end{itemize}

We have studied these quantities both numerically and analytically within the above model. We present below
some of our numerical results, and compare them with the empirical results of \cite{exp}. 
Our most important analytical result is the existence of a critical value $J_c$, below which the unpredictability
$U$ is strictly zero in the limit $N \to \infty$, and becomes positive for $J > J_c$, growing as $U \sim J - J_c$
close to the transition. The fluctuations of $U$ diverge close to $J_c$, as for standard second order phase
transitions. The value of $J_c$ can be computed exactly in the limit of a large number of items $M \gg 1$, and
depends on the detailed shape of the distribution of the fields $F$ and $h$. More precisely, $J_c$ is given by:
\be
J_c =  \int_{-\infty}^\infty {\rm d} F \, P_F(F) \gamma(F),
\ee
where $\gamma(F)$ is the solution of:
\be
\gamma = \int_{J_c-F-\gamma}^\infty {\rm d} u \, \frac{P_h(u)}{P_h(0)},
\ee
and $P_F$ and $P_h$ are the distributions of the fields $F$ and $h$.

\section{The Web-based experiment of Salganik et al.}

Here we describe the beautiful experimental set-up of 
M. J. Salganik, P. S. Dodds and D. J. Watts \cite{exp}, which allows them to
conclude that social influence has a determinant effect on the choices of 
individual agents. In the next section, we will in fact use their quantitative results to measure, 
within the above theoretical framework, the strength of the social influence factor $J$. 
Salganik et al. have \cite{exp} created an
artificial ``music market'' on the web with $M=48$ songs from 
essentially unknown bands in which 14,341 (mostly teen-agers) participated. Songs are
presented in a screen and participants make decisions about which
songs to listen to, and in a second step, whether they want to download the song they listened to. 
Participants are randomly assigned to one of the three following situations: 
\begin{itemize} 

\item an independent (zero-influence) situation where the list of songs carries no mention of the 
songs downloaded by other participants. This situation allows to define a benchmark,
where an `intrinsic' mix between the quality of the songs and the preference of the
participants can be measured. This situation corresponds to $J=0$ in the model above;

\item a `weak' social influence situation. In this case, the number of times a 
given song has been downloaded by other participants is shown. However, the songs are 
presented in random order so that the ranking of the preference of other participants
is not obvious at first glance. This situation corresponds to a certain small value $J_1  > 0$ 
in the model above;

\item a `strong' social influence situation. In this case, the list of songs is presented 
by decreasing number of downloads, such as to emphasize the  
preferences expressed by previous participants. This situation corresponds to a certain value $J_2 > J_1  > 0$ 
in the model above.

\end{itemize} 

Furthermore, in both
social influence conditions participants are randomly assigned to
$W=8$ different worlds, each one with its own history and evolving
independently from one another, but with the same initial conditions,
i.e. zero downloads. For each of the two influence conditions, the outcomes 
(i.e. the number of downloads of all songs) are compared to the independent, zero-influence situation.
In this way, the authors are able to conclude that increasing the strength of social influence increases both
the inequality $G$ and the unpredictability of success $U$ \cite{exp}.

Because these experiments look very much like those in physical
laboratories, we believe that they could play an important role
in the development of scientific investigations of collective human behavior. 
The Web gives the opportunity to devise and perform large scale experimentation (see also \cite{ZhangMG}),
with a number of participants that allows one to extract meaningful statistical information, 
We expect that many other experiments of the same type will be conducted in the future.
In the present case, the experiment is very carefully thought through to remove many artefacts: 
for example, download is free (no consideration of the wealth of participants is required -- no `budget constraint') 
and anonymous (no direct social pressure is involved);
participants are not rewarded to have made a `good' or `useful' choice, songs and bands are not well known (avoiding 
strong a priori biases), etc. 

\begin{figure}
\vskip 0.2cm
\begin{center}
\epsfig{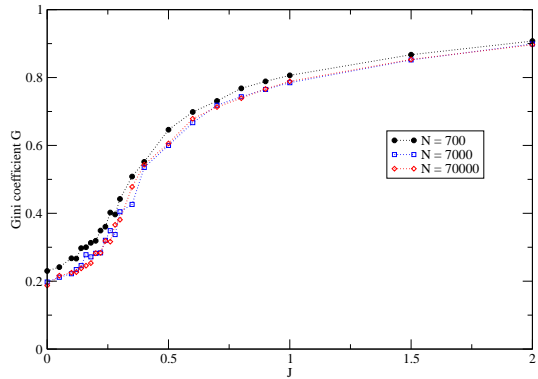}
\caption{
Gini coefficient as a function of $J$ for the choice $m_F\approx -2$, $\Sigma_F\approx 0.2$, $\Sigma=1$, and for different 
number of agents $N=700, 7000$ and $70000$. Note the rather weak dependence on $N$ of this quantity. The empirical
values of $G$ in the three different situations are: $G_0 \approx 0.22$ (no imitation),  $G_1 \approx 0.35$ (weak 
imitation) and  $G_2 \approx 0.5$ (strong imitation). 
}
\end{center}
\end{figure}

\section{Model calibration: towards a measurement of social influence?}

We now turn to a semi-quantitative analysis of the empirical data collected by Salganik et al. \cite{exp}. 
Once the distribution of $F^\alpha$'s and $\hia$'s are fixed (we chose them to be Gaussian for simplicity),
the model depends on four parameters: $m_F, \Sigma_F, \Sigma$ and the social influence $J$. These values 
must be chosen as to reproduce the observations reported in \cite{exp}, namely:
\begin{itemize}

\item The Gini coefficient $G_0$, the unpredictability $U_0$ and the qualitative shape of the distribution 
of $\phi^\alpha_0$ in the independent situation, corresponding to $J=0$. 

\item The Gini coefficient $G$, the unpredictability $U$ and the qualitative shape of the relation between
$\phi^\alpha$ and $\phi^\alpha_0$ in the social influence conditions

\end{itemize}

\begin{figure}
\vskip 0.5cm
\begin{center}
\epsfig{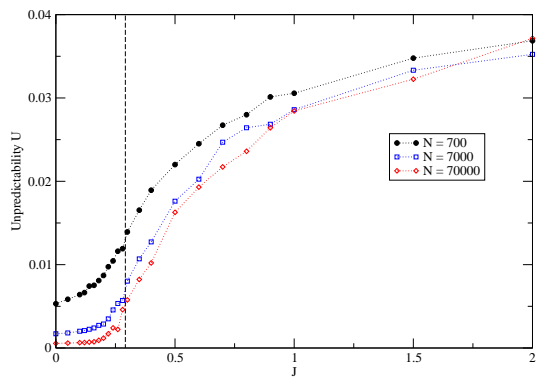}
\caption{
Unpredictability $U$ as a function of $J$ for the choice $m_F\approx -2$, $\Sigma_F\approx 0.2$, $\Sigma=1$, and for different 
number of agents $N=700, 7000$ and $70000$. In this case, the finite size effects are strong; one in fact expects 
$U$ to be zero for $J < J_c \approx 0.29$ (dashed vertical line), and to grow linearly for small $J - J_c >0$. 
The empirical values of $U$ for $N=700$ and 
in the three different situations are: $U_0 \approx 0.0045$ (no imitation),  $U_1 \approx 0.008$ (weak 
imitation) and  $G_2 \approx 0.013$ (strong imitation). This last case corresponds, for $N = 700$, to $J_2 \sim J_c$.
}
\end{center}
\end{figure}

Quite a lot more data is reported in the supplementary material of \cite{exp}, for example the average
number of downloaded songs per participant $d = {\cal C}/N$. In fact, the situation
of \cite{exp} is slightly more complicated than assumed in the above model because each participant makes a two-step
decision. Participants, before possibly downloading a song, first choose to listen to it. These two decisions 
may be correlated and both influenced by the choice of other participants. The authors of \cite{exp}  report 
separate statistics for the number of downloaded songs and the number of `tested' songs. In order to 
reproduce these results in full detail, one must generalize the above model, for example by assuming that the 
number of downloads of song $\alpha$ by agent $i$ can be written as:
\be
n_i^\alpha(t+1) = \Psi_i^\alpha \, \Theta\left[\, F^\alpha + {\overline h}_i + \delta h_i^\alpha + 
JM \left(\phi^\alpha(t) - \frac{1}{M} \right)\,\,\right],
\ee
where $\Psi_i^\alpha = 1$ with probability $p^\alpha$ and $0$ otherwise describing the decision of 
actually downloading a song after listening to it. Although the inclusion of this second decision step
is crucial to account fully for the results of \cite{exp}, we neglect this aspect altogether in the present paper and 
refer the reader to a later, more detailed publication \cite{ustocome}. Here we want to show that the main   
empirical features can indeed be reproduced by the model. 

Different choices of $m_F, \Sigma_F, \Sigma$ are in fact compatible with the observations corresponding to $J=0$, for which 
Salganik et al. find $G_0 \approx 0.22$ and $U_0 \approx 0.0045$ (for a number of participants in each `world' of $N=700$, 
the value we also use in our numerical simulations). A possible choice (further justified in \cite{ustocome}) 
is: $m_F\approx -2$, $\Sigma_F\approx 0.2$, $\Sigma=1$. The resulting shape of the distribution of $\phi^\alpha_0$ is found to 
be compatible with the data of \cite{exp}. Note that 
$\Sigma_F^2 = 0.04 < \Sigma^2 + \sigma^2 = 2$, suggesting that the intrinsic quality of songs is less dispersed than the 
preference of agents. This is expected in a situation where songs and bands are unknown, leading to very small
{\it a priori} information on their intrinsic quality.

Now, it is interesting to see how $G$ and $U$ are affected by a non zero value of $J$ -- cf. Figs. 1 and 2. 
>From these plots, one sees that the `weak' social influence situation, characterized by $G_1 \approx 0.35$ and 
$U_1 \approx 0.008$ \cite{exp}, corresponds to $J_1 \approx 0.17$. One the other hand, the `strong' influence situation 
yields $G_2 \approx 0.5$ and $U_2 \approx 0.013$ \cite{exp}, which we can account for by setting $J_2 \approx 0.30$. 
The scatter plots of $\phi^\alpha$ vs. $\phi^\alpha_0$ are shown in Figs 3-a and 3-b and can be satisfactorily 
compared to Figs. 3-A and 3-C of \cite{exp}. 

\begin{figure}
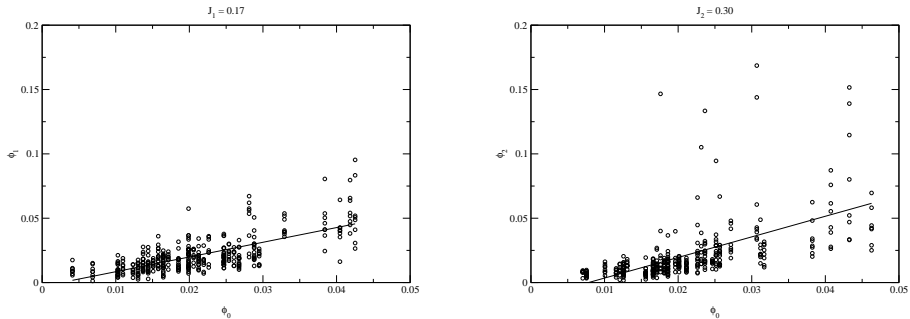

\vskip 0.2cm
\begin{center}
\epsfig{file=exp1-mu.eps,width=5.5cm}\hskip 1cm \epsfig{file=exp2-mu.eps,width=5.5cm}
\caption{
Scatter plot of the realized preferences $\phi^\alpha(J)$ as a function of the `intrinsic' preferences $\phi^\alpha_0$,
in the weak social influence condition ($J_1=0.17$, left), and in the strong social influence condition ($J_2=0.30$, right), all
for $m_F\approx -2$, $\Sigma_F\approx 0.2$, $\Sigma=1$.
Lines are linear regressions. 
These plots compare well with the corresponding plots of \cite{exp} (Figs. 3-A and 3-C; we use here the same scale as in
\cite{exp}).
}
\end{center}
\end{figure}

It is of particular interest to compare the above values of $J_1$ and $J_2$ to the critical value $J_c$ of the 
model, which can be determined exactly as a function of $m_F, \Sigma_F, \Sigma$ in the limit $M \to \infty$ \cite{ustocome}. 
In the present
case, we find $J_c \approx 0.29$, such that, in the limit $N \to \infty$, $U(J < J_c)$ should be strictly zero. 
As expected on general grounds and shown in Fig. 2, the value of $U$ at finite $N$ suffers from large finite size 
effects. Only a careful extrapolation for $N \to \infty$ allows one to confirm the existence of a critical value $J_c$
\cite{ustocome}. But in any case, the value $J_2$ accounting for the data in the `strong' influence situation is indeed 
quite large, since it corresponds to the critical region where imitation effects become dominant. 

Another effect worth noticing is the dependence of the average number of downloaded songs $d$ (or 
consumption ${\cal C} = Nd$) on the imitation parameter $J$, predicted by the model and reported in Fig 4. 
We see that this quantity has a clear {\it maximum}
as a function of $J$: at first, imitation effects tend to increase the total consumption until $J \sim 1$, beyond which
over-polarisation on a small number of items become such that the total consumption goes back down. This might 
have interesting consequences for marketing policies, for example (see e.g. \cite{Bass,Steyer}). The increase of the
$d$ with $J$ is actually not observed in \cite{exp}; see \cite{ustocome} for a further discussion of this point.

\section{Conclusions}

We have proposed a generic model for {\it multiple} choice situations with imitation effects and compared it with
recent empirical results from a Web-based cultural market experiment. Our model predicts to a 
phase transition between a weak imitation phase, in which expressed individual preferences are close to 
their value in the absence of any direct social pressure, and a strong imitation, `fashion' phase, where choices are 
driven by peer pressure and the ranking of individual preferences is strongly distorted at the aggregate level. 
The model can be calibrated to reproduce the main experimental results of Salganik et al. \cite{exp}; we show in particular that
the value of the social influence parameter can be estimated from the data. In one of the experimental situation, this 
value is found to be close to the critical value of the model, confirming quantitatively that social pressure are 
strong in that case. This concurs with the conclusions of \cite{QMB}, who also found near critical values of the social influence parameter.

Our model can be transposed to many interesting situations, for example that of industrial production, for which
one expects a transition between an archaic economy dominated by very few products and a fully diversified economy as the
dispersion of individual needs becomes larger. We leave the investigation of these questions, and the detailed 
analytical investigation of our model, for a further publication. We believe that the simultaneous development 
of theoretical models and detailed, rigorous experiments in the vein of \cite{exp} or \cite{ZhangMG,starflag}, will help promoting
a quantitative understanding of collective human (and animal) behaviour. 

\section*{Acknowledgments}

C.B. thanks Bertrand Roehner for useful conversations and for his enlightening efforts to extend 
physical intuition towards social sciences. J.P.B thanks Raphael Hitier for bringing ref. \cite{exp} to 
his attention. We also warmly thank Matteo Marsili for many discussions over the years, and for hospitality during 
completion of this work. 

\begin{figure}
\vskip 0.2cm
\begin{center}
\epsfig{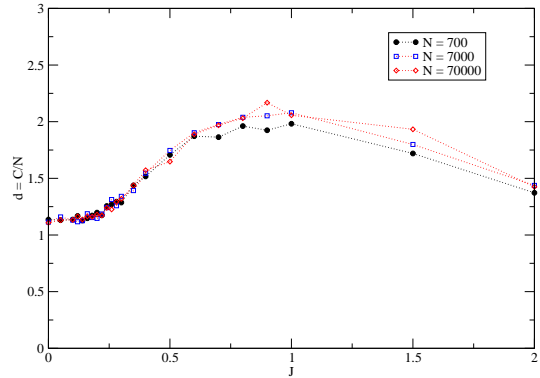}
\caption{
Average number of downloaded songs $d$ (or  consumption ${\cal C} = Nd$) as a function of $J$ for the choice 
$m_F\approx -2$, $\Sigma_F\approx 0.2$, $\Sigma=1$, and for different 
number of agents $N=700, 7000$ and $70000$. Finite size effects are quite small in this case. 
Note the clear maximum of this quantity as a function of the imitation 
strength $J$.
}
\end{center}
\end{figure}

\end{document}